\documentclass[a4paper]{report}
\usepackage[utf8]{inputenc}
\usepackage[T1]{fontenc}
\usepackage{RJournal}
\usepackage{amsmath,amssymb,array}
\usepackage{booktabs}

\usepackage{longtable}

\makeatletter
 \let\@cite@ofmt\@firstofone
 \def\@biblabel#1{}
 \def\@cite#1#2{{#1\if@tempswa , #2\fi}}
\makeatother
\newlength{\cslhangindent}
\newlength{\csllabelwidth}
 {\begin{list}{}{%
  \setlength{\itemindent}{0pt}
  \setlength{\leftmargin}{0pt}
  \setlength{\parsep}{0pt}
  \ifodd #1
   \setlength{\leftmargin}{\cslhangindent}
   \setlength{\itemindent}{-1\cslhangindent}
  \fi
  \setlength{\itemsep}{#2\baselineskip}}}
 {\end{list}}
\usepackage{calc}

\usepackage{floatrow}
\usepackage{multirow}
\usepackage{placeins}
\begin{document}

\sectionhead{}
\volume{}
\volnumber{}
\year{}
\month{}
\fancyhf{}
\fancyfoot[C]{\thepage}
\renewcommand{\headrulewidth}{0pt}
\hypersetup{pdftitle={ShrinkageTrees: An R Package for Bayesian Tree Ensembles
  for Survival Analysis and Causal Inference}}

\begin{article}
\title{ShrinkageTrees: An R Package for Bayesian Tree Ensembles for Survival Analysis and Causal Inference}

\author{by Tijn Jacobs}

\maketitle

\vspace{-0.6\baselineskip}%
\noindent\emph{Department of Mathematics, Vrije Universiteit Amsterdam, Amsterdam, The Netherlands\\%
\href{mailto:t.jacobs@vu.nl}{\nolinkurl{t.jacobs@vu.nl}}}

\vspace{1.3\baselineskip}

\abstract{%
ShrinkageTrees is an R package for Bayesian tree ensembles in survival analysis and causal inference. The package implements Bayesian additive regression tree models for right- and interval-censored survival outcomes within an accelerated failure time (AFT) framework, with optional decomposition into prognostic and treatment-effect components for causal inference. Two complementary forms of regularisation are available: regularisation of the tree structure, via depth-penalising priors and Dirichlet splitting priors, and regularisation of the step heights, via global--local shrinkage priors. ShrinkageTrees provides the first implementation of the Horseshoe Forest, which places a horseshoe prior on the step heights. These regularisation strategies extend Bayesian tree ensembles to high-dimensional settings. An efficient Rcpp backend, multi-chain MCMC, and S3 methods support the full workflow: fitting, prediction, causal effect estimation, and convergence diagnostics.
}

\section{Introduction}\label{introduction}

Survival analysis in high-dimensional settings poses three interrelated challenges. First, the relationship between covariates and the time-to-event outcome is often complex and unknown, calling for flexible non-parametric models. Second, the outcome may be observed imprecisely: right-censored when a patient is lost to follow-up, or interval-censored when the event is known only to have occurred between two assessment times. Third, when a treatment is involved, even the average causal effect is challenging to estimate under censoring and confounding. The effect may also vary across patient subgroups, and disentangling this heterogeneity from baseline patient characteristics requires dedicated methodology. These challenges arise routinely in clinical trials, observational studies, and genomics, often simultaneously.

Bayesian Additive Regression Trees {[}BART; \citet{chipman2010bart}{]} offer a natural starting point. BART models the response as a sum of many shallow decision trees, where each tree contributes a small piece of the overall prediction. The ensemble captures non-linearities and interactions without requiring a parametric form. Existing BART implementations leave gaps in each of the three areas above. Currently, no R package provides Bayesian tree ensembles for interval-censored survival outcomes. The \(\tau\)-learner is a dedicated causal forest architecture that separates the prognostic function from the treatment effect \citep{hahn2020bayesian, caron2022estimating}. It has not been implemented for time-to-event endpoints. Standard BART's regularisation, while adequate in moderate dimensions, does not adapt well to the high-dimensional settings (\(p \gg n\)) common in genomics.

\CRANpkg{ShrinkageTrees} addresses all three gaps. The package fits Bayesian tree ensembles to right- and interval-censored survival data via accelerated failure time (AFT) models, with optional \(\tau\)-learner decomposition for causal inference. The package implements two complementary regularisation strategies for high-dimensional settings. Regularisation of the tree structure encourages the ensemble to split on a subset of relevant variables. The package implements the Dirichlet splitting approach of DART \citep{linero2018dart}. Regularisation of the step heights targets the contributions of individual leaves: it shrinks uninformative leaves towards zero while preserving strong signals. ShrinkageTrees provides the first implementation of the Horseshoe Forest \citep{jacobs2025horseshoe}, which places a horseshoe prior \citep{carvalho2010horseshoe} on the step heights to adaptively shrink uninformative leaves.

The package is implemented with an efficient C++ backend via \CRANpkg{Rcpp} \citep{eddelbuettel2011rcpp} and provides a coherent S3 interface with \texttt{print()}, \texttt{summary()}, \texttt{predict()}, and \texttt{plot()} methods. The paper is aimed at both practitioners encountering Bayesian tree ensembles for the first time and experienced users seeking functionality not available in existing packages.

To give a first impression, we use the \texttt{ovarian} dataset shipped with the package: a semi-synthetic cohort with covariates drawn from The Cancer Genome Atlas ovarian cancer cohort {[}TCGA-OV; \citet{tcga2011ovarian}{]}. It comprises 357 patients with right-censored overall survival, a binary treatment indicator, four clinical covariates, and expression levels of 1,000 genes. This is a high-dimensional (\(p \gg n\)) setting that exercises the main features of the package. A Horseshoe Forest can then be fitted in a single call:

\begin{verbatim}
library(ShrinkageTrees)
data("ovarian")
time   <- ovarian$OS_time / 30.44  # days to months
status <- ovarian$OS_event
X      <- as.matrix(ovarian[, -(1:3)])
fit    <- HorseTrees(y = time, status = status, X_train = X,
                     outcome_type = "right-censored", number_of_trees = 200)
\end{verbatim}

The returned object is of class \texttt{ShrinkageTrees}. It supports \texttt{print()}, \texttt{summary()}, \texttt{predict()}, and \texttt{plot()} methods for inspection, prediction, and posterior survival curves.

\section{A worked example: Survival prediction on the ovarian dataset}\label{a-worked-example-survival-prediction-on-the-ovarian-dataset}

We introduce the package through a complete worked example before discussing the underlying methodology. We return to the \texttt{ovarian} dataset loaded in the introduction. The treatment indicator distinguishes carboplatin from cisplatin, and the four clinical covariates are age, FIGO stage, tumour grade, and year of diagnosis. The 1,000 gene expression features are selected by median absolute deviation. Ovarian cancer is one of the most lethal gynaecological malignancies, with a five-year survival rate below 50\%, and gene expression patterns are known to be associated with survival duration and chemotherapy response \citep{tcga2011ovarian}. We additionally extract the treatment indicator, used in Section 4.2.

\begin{verbatim}
treatment <- ovarian$treatment
\end{verbatim}

We begin with the simplest model available in \CRANpkg{ShrinkageTrees}: a standard BART survival model via \texttt{SurvivalBART()}. This uses an accelerated failure time (AFT) formulation with right-censoring and a conjugate normal prior on the step heights. It is the classical BART specification of \citet{chipman2010bart} applied to survival data. We fit the model on the full cohort.

\begin{verbatim}
fit_bart <- SurvivalBART(
  time            = time,
  status          = status,
  X_train         = X,
  timescale       = "time",
  number_of_trees = 200,
  N_post          = 5000,
  N_burn          = 5000,
  n_chains        = 4
)
\end{verbatim}

The \texttt{print()} method provides a concise overview of the fitted model.

\begin{verbatim}
print(fit_bart)
\end{verbatim}

\begin{verbatim}

ShrinkageTrees model
---------------------
Outcome type:         Right-censored (AFT, timescale = time)
Prior:                standard
Number of trees:      200
Training size (n):    357
Number of features:   1004
Chains:               4
Draws per chain:      5000 (burn-in 5000)
Acceptance ratio:     0.243, 0.243, 0.246, 0.246 (per chain)
Posterior mean sigma: 0.23
\end{verbatim}

We evaluate in-sample fit using the concordance index {[}C-index; \citet{harrell1996concordance}{]} via the \CRANpkg{survival} package \citep{therneau2024survival}, where a value of 1 indicates perfect discrimination and 0.5 corresponds to random prediction.

\begin{verbatim}
library(survival)
c_train <- concordance(Surv(time, status) ~ fit_bart$train_predictions)
cat("Train C-index:", round(c_train$concordance, 3), "\n")
\end{verbatim}

\begin{verbatim}
Train C-index (SurvivalBART): 0.989
\end{verbatim}

The training C-index is close to one: standard BART discriminates the observed events almost perfectly. In a \(p \gg n\) regime with \(n = 357\) and \(p = 1{,}004\), such near-perfect in-sample fit is a red flag. The simulation in Section 4.3 confirms this on data with known ground truth: standard BART recovers the signal in point estimates but its credible intervals are systematically miscalibrated. Stronger regularisation is needed. \CRANpkg{ShrinkageTrees} offers several options, including a Dirichlet splitting prior on the tree structure {[}DART; \citet{linero2018dart}{]} and global--local shrinkage priors on the step heights \citep{jacobs2025horseshoe}, of which the horseshoe is the headline contribution. We introduce these options in Section 3 and revisit the ovarian example with horseshoe shrinkage in Section 4.1.

The \texttt{plot()} method produces posterior survival curves. The population-averaged curve across all observations can be overlaid with the Kaplan--Meier estimate for visual calibration:

\begin{verbatim}
plot(fit_bart, type = "survival", km = TRUE)
\end{verbatim}

\begin{figure}

{\centering \includegraphics[width=0.65\linewidth]{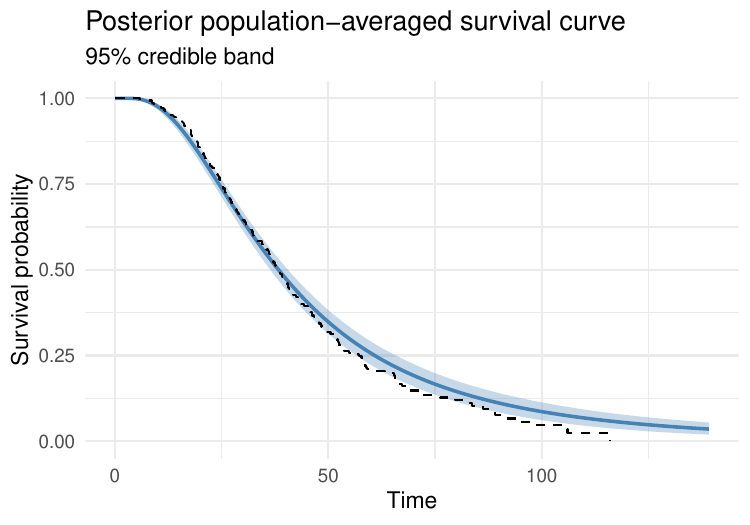} 

}

\caption[Population posterior survival curve (BART baseline).]{Population-averaged posterior survival curve under the standard BART baseline (solid line) with 95\% credible band (shaded) and Kaplan--Meier estimate (dashed).}\label{fig:survival-bart-population-fig}
\end{figure}

Figure \ref{fig:survival-bart-population-fig} shows the population-averaged curve tracking the Kaplan--Meier estimate closely. We first review the statistical methodology that underpins the extensions used in Section 4, where we progressively add horseshoe shrinkage, causal inference, and interval-censored outcomes.

\section{Statistical methodology}\label{statistical-methodology}

We formalise the models behind \CRANpkg{ShrinkageTrees}. First, we review the BART model and the AFT formulation for survival outcomes. This is the model fitted by \texttt{SurvivalBART()} in Section 2. Second, we show how regularisation targets either the tree structures or the step heights. Third, we extend the framework to causal inference via the \(\tau\)-learner.

We use the following notation throughout. The data consist of \(n\) subjects indexed by \(i = 1, \ldots, n\), each with a covariate vector \(\mathbf{x}_i \in \mathbb{R}^p\) and a response \(Y_i\) (continuous in the BART subsection, \(\log T_i\) in the AFT subsection). An ensemble consists of \(m\) trees indexed by \(j = 1, \ldots, m\). Bold lowercase letters denote vectors and calligraphic letters denote tree-related objects: \(\mathcal{T}_j\) for the tree structure and \(\mathcal{H}_j\) for its step heights.

\subsection{BART and the accelerated failure time model}\label{bart-and-the-accelerated-failure-time-model}

\begin{figure}

{\centering \includegraphics[width=0.85\linewidth]{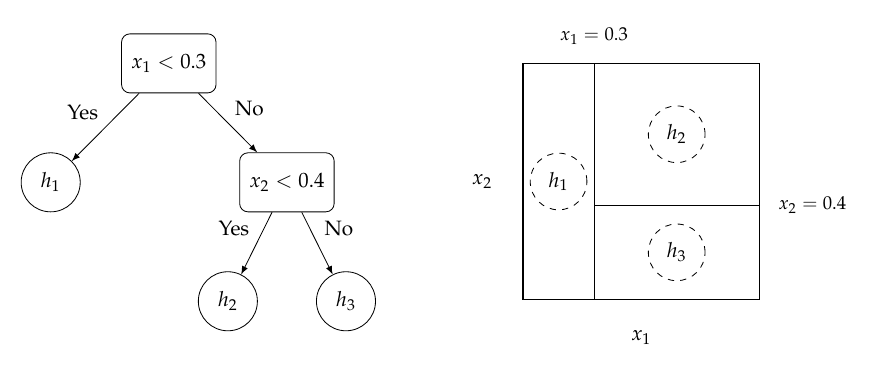} 

}

\caption[Schematic of a regression tree.]{Schematic of a single regression tree. Interior nodes contain binary splitting rules of the form $x_\rho < c$; terminal nodes (leaves) contain step heights $h_\ell$. An observation traverses the tree from root to leaf, and its prediction is the step height of the leaf it reaches.}\label{fig:ExampleTree}
\end{figure}

BART \citep{chipman2010bart} models the response as a sum of \(m\) shallow decision trees:
\begin{equation}
Y_i = \sum_{j=1}^{m} g(\mathbf{x}_i;\, \mathcal{T}_j, \mathcal{H}_j) + \varepsilon_i, \quad \varepsilon_i \sim \mathcal{N}(0, \sigma^2).
\label{eq:bart}
\end{equation}
Each tree \(j\) is characterised by two groups of parameters. The \emph{tree structure} \(\mathcal{T}_j\) encodes the topology and the splitting rules: which variables to split on and at what thresholds. It determines how the covariate space is partitioned into disjoint regions. The \emph{step heights} \(\mathcal{H}_j = \{h_{j1}, \ldots, h_{jL_j}\}\) are the values assigned to the \(L_j\) leaves and determine each region's contribution to the fitted function. For an input \(\mathbf{x}\), the function \(g(\mathbf{x};\, \mathcal{T}_j, \mathcal{H}_j)\) returns the step height of the leaf that \(\mathbf{x}\) reaches (see Figure \ref{fig:ExampleTree}). The distinction between \(\mathcal{T}\) and \(\mathcal{H}\) is central to the regularisation framework we develop below.

We specify priors that are independent across trees and factorise within each tree:
\begin{equation}
p(\{\mathcal{T}_j, \mathcal{H}_j\}_{j=1}^m) = \prod_{j=1}^{m} p_{\mathcal{T}}(\mathcal{T}_j)\, p_{\mathcal{H}}(\mathcal{H}_j \mid \mathcal{T}_j).
\label{eq:priors}
\end{equation}
The tree structure prior \(p_{\mathcal{T}}\) is a branching process \citep{chipman1998bayesiancart, rockova2020theory} in which a node at depth \(d\) splits with probability \(\alpha\, (1 + d)^{-\beta}\), with \(\alpha \in (0, 1)\) and \(\beta \geq 0\). The default values \(\alpha = 0.95\) and \(\beta = 2\) keep most trees shallow. At each internal node, the splitting variable is drawn uniformly from the \(p\) covariates and the split point uniformly from the observed range of that variable.

The original BART formulation \(p_{\mathcal{H}}\) assigns Gaussian priors to the step heights, independent across leaves and trees, with variance \(\sigma_h^2 = 1/(4 k^2 m)\) on the centred and scaled response. The default \(k = 2\) places approximately 95\% of the prior mass within the observed range, following \citet{chipman2010bart}.

We model the error variance as \(\sigma^2 \sim \text{Inv-}\chi^2(\nu,\, \psi)\), with \(\nu = 3\) and \(\psi\) chosen so that the prior is centred around a rough estimate of \(\sigma^2\) \citep{chipman2010bart}.

For survival outcomes, we adopt an accelerated failure time (AFT) formulation. Let \(T_i\) denote the event time, \(C_i\) the censoring time, \(Y_i = \min(T_i, C_i)\) the observed follow-up time, and \(\delta_i \in \{0, 1\}\) the event indicator. The prediction model becomes:
\begin{equation}
\log T_i = \sum_{j=1}^{m} g(\mathbf{x}_i;\, \mathcal{T}_j, \mathcal{H}_j) + \varepsilon_i, \quad \varepsilon_i \sim \mathcal{N}(0, \sigma^2).
\label{eq:aft}
\end{equation}
For right-censored observations (\(\delta_i = 0\)), we augment \(\log T_i\) from a truncated normal distribution within the Gibbs sampler. \CRANpkg{ShrinkageTrees} also supports interval-censored outcomes, where the event time is known only to lie in an interval \((l_i, r_i]\). We augment these analogously. This is the model behind the \texttt{SurvivalBART()} call in Section 2.

\subsection{Regularisation}\label{regularisation}

The parameters of a BART ensemble fall into two groups: tree structure parameters (nodes, splitting variables, split points) that determine how the covariate space is partitioned, and step heights that determine each partition's contribution to the fitted function. Regularisation can target either group, or both.

\textbf{Tree structure regularisation.} The branching process prior on \(\mathcal{T}_j\) is itself a regularisation mechanism: through \(\alpha\) and \(\beta\), it discourages deep trees. DART \citep{linero2018dart} adds a second mechanism. Standard BART draws splitting variables uniformly from the \(p\) covariates. DART instead places a Dirichlet prior on the splitting probabilities \(\mathbf{s} = (s_1, \ldots, s_p)\):
\begin{equation}
\mathbf{s} \sim \text{Dirichlet}\!\left(\frac{\theta}{p}, \ldots, \frac{\theta}{p}\right).
\label{eq:dirichlet}
\end{equation}
Smaller values of \(\theta\) concentrate the prior on sparse probability vectors, so the ensemble splits on a subset of relevant variables. This is a form of structural variable selection. We follow the hyperprior calibration of \citet{linero2018dart} and place a hyperprior on \(\theta\):
\begin{equation}
\frac{\theta}{\theta + \rho} \sim \text{Beta}(a, b),
\label{eq:theta-hyper}
\end{equation}
with defaults \(a = 0.5\), \(b = 1\), and \(\rho = p\).

\textbf{Shrinkage on step heights.} Regularisation can also target the step heights directly. \CRANpkg{ShrinkageTrees} adopts a scale mixture of normals representation:
\begin{equation}
h_\ell \mid \lambda_\ell, \tau, \omega \sim \mathcal{N}\left(0,\, \omega\, \lambda_\ell^2\, \tau^2\right),
\label{eq:scale-mixture}
\end{equation}
where \(\tau\) is a global shrinkage parameter shared across all leaves in a tree, \(\lambda_\ell\) is a local scale specific to leaf \(\ell\), and \(\omega\) is a fixed scaling constant. The global parameter \(\tau\) controls the overall degree of shrinkage, while the local parameters \(\lambda_\ell\) allow individual leaves with strong signals to escape shrinkage.

The horseshoe prior \citep{carvalho2010horseshoe} is a specific instance of this framework, with independent half-Cauchy priors on the local and global scales:
\begin{equation}
\lambda_\ell \sim \text{C}^+(0, \alpha_\lambda), \qquad \tau \sim \text{C}^+(0, \alpha_\tau),
\label{eq:horseshoe}
\end{equation}
where \(\text{C}^+(0, \alpha)\) denotes the half-Cauchy distribution with scale \(\alpha\). The half-Cauchy has its median at its scale parameter and heavy tails. Its mean and variance do not exist, which allows the local scales to take very large values when the data support a strong effect \citep{polson2012halfcauchy}. The combination shrinks the contributions of uninformative leaves towards zero while preserving large, data-supported signals.

We set \(\alpha_\tau = \alpha_\lambda = k / \sqrt{m}\), following \citet{jacobs2025horseshoe}. Smaller values of \(k\) yield stronger shrinkage. We recommend \(k \in [0.05,\, 0.5]\) as a practical range. The default in \CRANpkg{ShrinkageTrees} is \(k = 0.1\), but the user can select \(k\) by cross-validation when domain knowledge does not suggest a specific value. \citet{jacobs2025horseshoe} provide a sensitivity analysis to support these recommendations.

\begin{figure}

{\centering \includegraphics[width=0.96\linewidth]{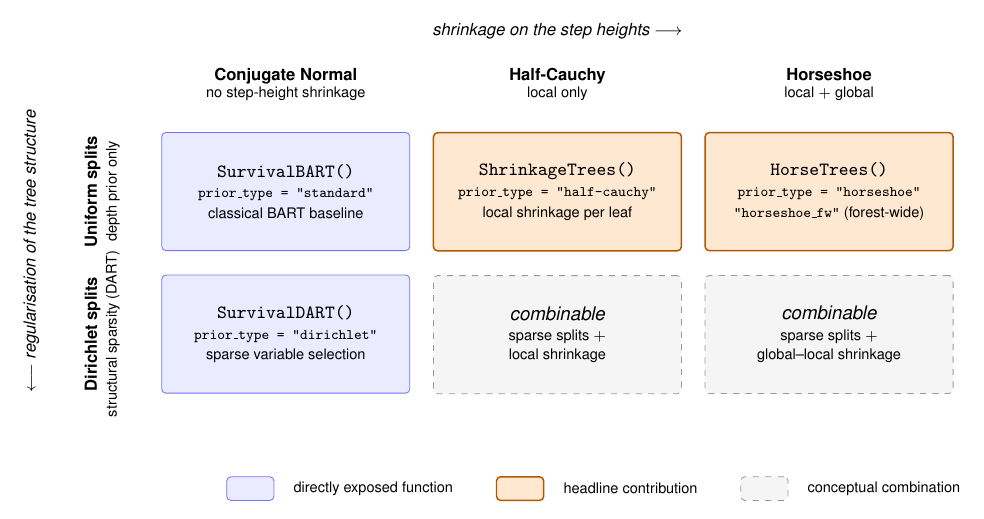} 

}

\caption[Regularisation landscape.]{Regularisation landscape for \CRANpkg{ShrinkageTrees}. Solid cells name the fitting function and \texttt{prior\_type} value exposed in the package. Dashed cells mark combinations within the framework that the current release does not expose.}\label{fig:regularisation-landscape}
\end{figure}

Figure \ref{fig:regularisation-landscape} maps the resulting design space. Four cells correspond to fitting functions exposed by the package, with \texttt{HorseTrees()} as the headline contribution. The two dashed cells combine Dirichlet splitting with local or global--local shrinkage on the step heights. These combinations are within the framework but are not yet exposed in the current release.

\FloatBarrier

\subsection{\texorpdfstring{The \(\tau\)-learner for causal inference}{The \textbackslash tau-learner for causal inference}}\label{the-tau-learner-for-causal-inference}

BART has become a popular tool for causal inference \citep{hill2011bayesian, dorie2019automated}. The BART formulation in equation \eqref{eq:bart} estimates a single function of the covariates, which we call the \emph{prediction model}. Prediction models support two simple causal-inference workflows: an S-learner includes the treatment indicator as a covariate, and a T-learner fits separate models per treatment arm \citep{caron2022estimating}. \CRANpkg{ShrinkageTrees} takes a different approach: the \(\tau\)-learner. This is the decomposition proposed by \citet{hahn2020bayesian} as Bayesian Causal Forests (BCF) and later termed the \(\tau\)-learner by \citet{caron2022estimating}. The \CRANpkg{bcf} and \CRANpkg{stochtree} packages implement the \(\tau\)-learner for continuous outcomes. \CRANpkg{ShrinkageTrees} extends it to survival endpoints. We write \(A_i \in \{0, 1\}\) for the binary treatment indicator and \(\hat{e}(\mathbf{x}_i) = \hat{P}(A_i = 1 \mid \mathbf{x}_i)\) for the estimated propensity score. The model is:
\begin{equation}
y_i = \mu(\mathbf{x}_i, \hat{e}(\mathbf{x}_i)) + A_i\, \tau(\mathbf{x}_i) + \varepsilon_i,
\label{eq:tau-learner}
\end{equation}
where \(\mu\) is a prognostic function modelled by a forest of \(m_\mu\) trees and \(\tau\) is a treatment effect function modelled by a separate forest of \(m_\tau\) trees. This separation allows the prognostic and treatment effect components to have different regularisation and complexity. The conditional average treatment effect is \(\text{CATE}(\mathbf{x}) = \tau(\mathbf{x})\), and the average treatment effect is \(\text{ATE} = \mathbb{E}[\tau(\mathbf{X})]\).

\begin{figure}

{\centering \includegraphics[width=0.95\linewidth]{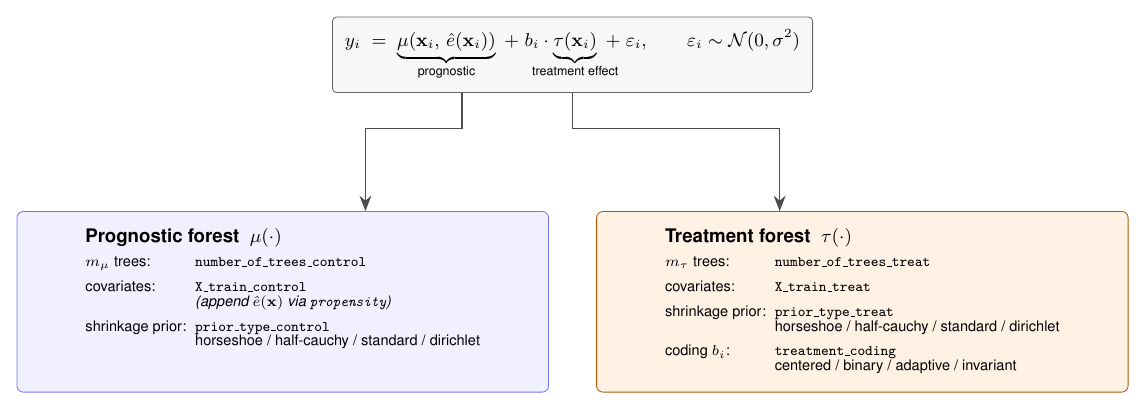} 

}

\caption[$\tau$-learner decomposition.]{The $\tau$-learner decomposition in \CRANpkg{ShrinkageTrees}. The outcome splits into a prognostic forest $\mu(\mathbf{x}, \hat e(\mathbf{x}))$ (blue) and a treatment-effect forest $\tau(\mathbf{x})$ (orange). Each box lists the function arguments that control the forest. For survival outcomes, set $y_i = \log T_i$.}\label{fig:tau-learner}
\end{figure}

The way treatment is coded in the \(\tau\)-learner is a modelling convention with several common choices. Let \(b_i\) denote the coding applied to subject \(i\). The model becomes:
\begin{equation}
y_i = \mu(\mathbf{x}_i, \hat{e}(\mathbf{x}_i)) + b_i\, \tau(\mathbf{x}_i) + \varepsilon_i.
\label{eq:tau-coding}
\end{equation}
\citet{hahn2020bayesian} discuss several choices, all of which are implemented in \CRANpkg{ShrinkageTrees} through the \texttt{treatment\_coding} argument. \textbf{Centred} coding sets \(b_i = A_i - 1/2\) and is the default. \textbf{Binary} coding sets \(b_i = A_i\). \textbf{Adaptive} coding sets \(b_i = A_i - \hat{e}(\mathbf{x}_i)\), the parameterisation used in the \CRANpkg{bcf} package. The \textbf{invariant} (parameter-expanded) coding of \citet[Section 5.2]{hahn2020bayesian} is invariant to the labelling of treatment and control. It reparameterises the model with group-specific intercepts and recovers \(\tau\) as a difference. We refer the reader to \citet{hahn2020bayesian} for the full construction. Figure \ref{fig:tau-learner} summarises the decomposition.

\subsection{Posterior computation}\label{posterior-computation}

All models are fitted using Markov chain Monte Carlo (MCMC). The outer loop is a Gibbs sampler that cycles through updates of the tree topologies, the step heights, and the error variance \(\sigma^2\). When applicable, it also updates the shrinkage parameters and censored event times. Trees are updated one at a time via Bayesian backfitting \citep{hastie2000backfitting}: at each iteration, a partial residual is formed by subtracting the contributions of all other trees from the response, and tree \(j\) is then updated conditional on the current state of all other trees. The tree topology is modified through Metropolis--Hastings proposals that either grow a terminal node into an internal node with two new leaves, or prune an internal node back to a leaf. Details of the sampler are given in \citet{kapelner2016bartmachine}.

Under the standard BART and Dirichlet (DART) priors, where the step heights have a conjugate normal prior, the marginal likelihood of a proposed tree can be computed in closed form and the step heights are integrated out analytically. When global--local shrinkage priors such as the horseshoe are placed on the step heights, this conjugacy is lost: the local and global scale parameters \(\lambda_\ell\) and \(\tau\) must be sampled explicitly, and the tree proposals become reversible jump MCMC moves. The details of this reversible jump construction are given in \citet{jacobs2025horseshoe}.

\section{A closer look at the ovarian example}\label{a-closer-look-at-the-ovarian-example}

We return to the ovarian cancer dataset from Section 2 and progressively demonstrate more advanced features of \CRANpkg{ShrinkageTrees}: horseshoe shrinkage for improved regularisation, causal inference for treatment effect estimation, and interval-censored survival modelling.

\subsection{Horseshoe shrinkage}\label{horseshoe-shrinkage}

We now fit a Horseshoe Forest via \texttt{HorseTrees()} to the same data. The horseshoe prior applies global--local shrinkage to the step heights. This provides stronger regularisation than the conjugate normal prior used by \texttt{SurvivalBART()}. It is particularly beneficial in this high-dimensional setting where the number of predictors far exceeds the sample size.

\begin{verbatim}
fit_horse <- HorseTrees(
  y               = time,
  status          = status,
  X_train         = X,
  outcome_type    = "right-censored",
  timescale       = "time",
  number_of_trees = 200,
  k               = 0.1,
  N_post          = 5000,
  N_burn          = 5000,
  n_chains        = 4
)
\end{verbatim}

The \texttt{print()} method returns the same configuration overview as in Section 2, now reporting the per-chain acceptance ratios and posterior mean of \(\sigma\) for the horseshoe fit. A richer \texttt{summary()} method adds credible intervals, prediction summaries, and convergence diagnostics from \CRANpkg{coda}; we demonstrate it on the causal model in Section 4.2.

We report the in-sample C-index for the horseshoe fit alongside the BART value from Section 2 for comparison. The first line of the output is repeated from Section 2. The second is new for the horseshoe fit.

\begin{verbatim}
c_train <- concordance(Surv(time, status) ~ fit_horse$train_predictions)
cat("Train C-index:", round(c_train$concordance, 3), "\n")
\end{verbatim}

\begin{verbatim}
Train C-index (SurvivalBART): 0.989
Train C-index (HorseTrees):   0.727
\end{verbatim}

The horseshoe produces a lower training C-index than \texttt{SurvivalBART()}. Standard BART concentrates ensemble capacity on whichever of the 1,004 covariates align with training events. The horseshoe shrinks contributions from leaves that are not strongly supported by the data. The simulation in Section 4.3 evaluates the two approaches on a controlled sparse setting where ground truth is available.

We assess convergence using \(\sigma\) traceplots and posterior densities. With \texttt{n\_chains\ =\ 4}, each chain is plotted separately, allowing visual assessment of mixing and chain agreement. The \texttt{summary()} method automatically reports the effective sample size and Gelman--Rubin \(\hat{R}\) statistic when the \CRANpkg{coda} package is installed.

\begin{verbatim}
plot(fit_horse, type = "trace")
plot(fit_horse, type = "density")
\end{verbatim}

\begin{figure}

{\centering \includegraphics[width=0.48\linewidth]{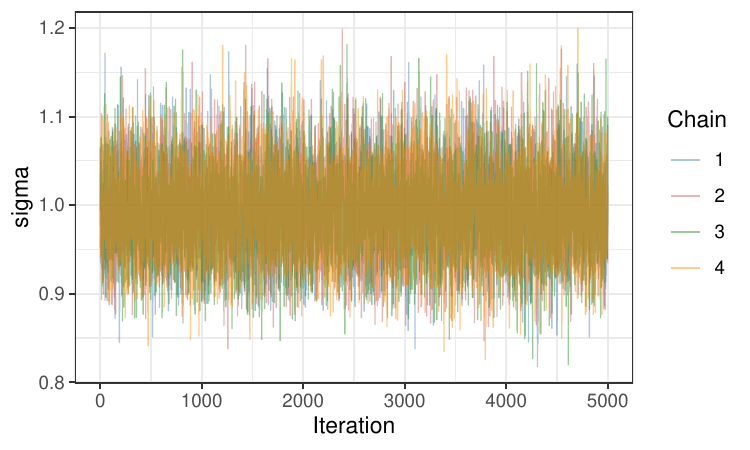} \includegraphics[width=0.48\linewidth]{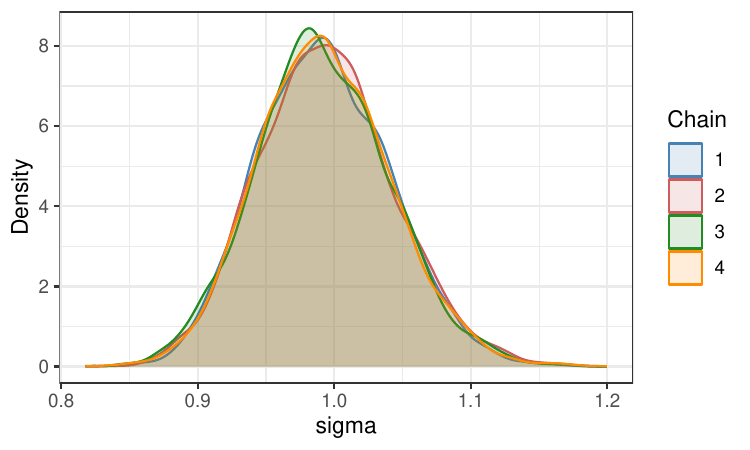} 

}

\caption[Convergence diagnostics.]{Left: traceplot of the posterior draws of $\sigma$ for the Horseshoe Forest model. Right: posterior density of $\sigma$, estimated separately for each of the four chains.}\label{fig:convergence-fig}
\end{figure}

Figure \ref{fig:convergence-fig} shows good mixing: the four chains overlap throughout the sampling period and yield nearly identical posterior densities for \(\sigma\).

\FloatBarrier

We illustrate the posterior predictive survival curve for a randomly selected patient.

\begin{verbatim}
pred <- predict(fit_horse, newdata = X)
idx <- sample(length(pred$mean), 1)
\end{verbatim}

The individual posterior survival curve under the AFT log-normal model \(S(t \mid \mathbf{x}) = 1 - \Phi((\log t - \hat{\mu}(\mathbf{x})) / \hat{\sigma})\) is plotted with pointwise 95\% credible bands.

\begin{verbatim}
plot(pred, type = "survival", obs = idx)
\end{verbatim}

\begin{figure}

{\centering \includegraphics[width=0.65\linewidth]{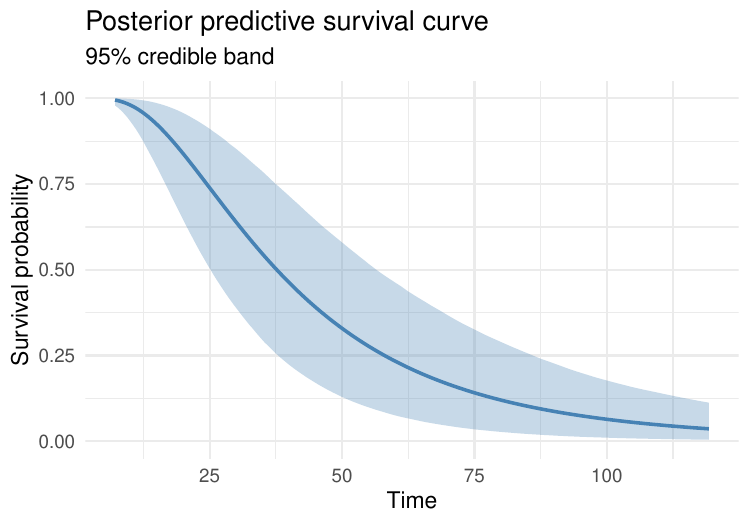} 

}

\caption[Posterior survival curve (Horseshoe Forest).]{Posterior survival curve for a randomly selected patient under the Horseshoe Forest, with pointwise 95\% credible bands.}\label{fig:survival-fig}
\end{figure}

Figure \ref{fig:survival-fig} shows the posterior survival curve for the selected patient with full uncertainty quantification.

\subsection{Causal inference: treatment effects for survival}\label{causal-inference-treatment-effects-for-survival}

We now estimate heterogeneous treatment effects of carboplatin versus cisplatin using \texttt{CausalHorseForest()}, which fits a \(\tau\)-learner with horseshoe shrinkage on both the prognostic and treatment effect forests. We first estimate propensity scores using a probit Horseshoe Forest on the full covariate matrix. We choose a Horseshoe Forest over logistic regression because it handles potential non-linearities and interactions without manual specification, and the horseshoe prior provides the regularisation needed in the \(p \gg n\) regime.

\begin{verbatim}
ps_fit <- HorseTrees(
  y               = treatment,
  X_train         = X,
  outcome_type    = "binary",
  number_of_trees = 200,
  k               = 0.1,
  N_post          = 5000,
  N_burn          = 5000
)
propensity <- ps_fit$train_predictions
X_control  <- cbind(propensity = propensity, X)
\end{verbatim}

\begin{verbatim}
fit_causal <- CausalHorseForest(
  y                         = log(time),
  status                    = status,
  X_train_control           = X_control,
  X_train_treat             = X,
  treatment_indicator_train = treatment,
  outcome_type              = "right-censored",
  timescale                 = "log",
  number_of_trees           = 200,
  k                         = 0.1,
  N_post                    = 5000,
  N_burn                    = 5000,
  n_chains                  = 4
)
summary(fit_causal)
\end{verbatim}

\begin{verbatim}

CausalShrinkageForest model summary
=====================================
Call: CausalHorseForest(y = log(time), status = status, X_train_control = X_control, 
    X_train_treat = X, treatment_indicator_train = treatment, 
    outcome_type = "right-censored", timescale = "log", number_of_trees = 200, 
    k = 0.1, N_post = 5000, N_burn = 5000, store_posterior_sample = TRUE, 
    n_chains = 4, verbose = TRUE)

Outcome: Right-censored (AFT, timescale = log)
Prior:   control = horseshoe, treatment = horseshoe
Trees:   control = 200, treatment = 200
Data:    n = 357, p_control = 1005, p_treat = 1004 | Draws: 5000 x 4 chains (burn-in 5000)

Treatment effect:
  PATE:    0.0207  95% CI (Bayesian bootstrap): [-0.1, 0.1474]
  CATE SD: 0.0077

Prognostic function (mu):
  Mean: 3.63  SD: 0.01  Range: [3.605, 3.648]

Posterior sigma:
  Mean: 1  SD: 0.048  95% CI: [0.913, 1.099]

MCMC acceptance ratios (per chain):
  control:   0.508, 0.507, 0.508, 0.508
  treatment: 0.475, 0.474, 0.475, 0.475
\end{verbatim}

The \texttt{summary()} method reports the posterior average treatment effect (ATE) on the log-survival scale with a 95\% credible interval. Here we pass \texttt{log(time)} directly with \texttt{timescale\ =\ "log"}, so all treatment effect estimates remain on the log-survival scale. A positive ATE would indicate that carboplatin is associated with longer survival times. By default, the reported interval is the population ATE {[}PATE; \citet{li2023bayesian}{]} obtained by a Bayesian bootstrap \citep{rubin1981bayesian}. At each MCMC iteration, the per-observation CATEs are reweighted with \(\text{Dirichlet}(1, \ldots, 1)\) weights before averaging. The credible interval then propagates uncertainty in the covariate distribution \(F_X\) rather than conditioning on the sample. The argument \texttt{bayesian\_bootstrap\ =\ FALSE} recovers the narrower mixed ATE {[}MATE; \citet{li2023bayesian}{]}, which conditions on the observed covariates and therefore tends to be overconfident. The output also reports the prognostic function \(\mu\), the residual scale \(\sigma\), and per-chain acceptance ratios as diagnostic information.

The \texttt{plot()} method provides visualisations of the treatment effect posterior. The ATE density plot shows the full posterior distribution of the average treatment effect, while the CATE caterpillar plot displays patient-level treatment effect estimates sorted by posterior mean with 95\% credible intervals.

\begin{verbatim}
plot(fit_causal, type = "ate")
plot(fit_causal, type = "cate")
\end{verbatim}

\begin{figure}

{\centering \includegraphics[width=0.48\linewidth]{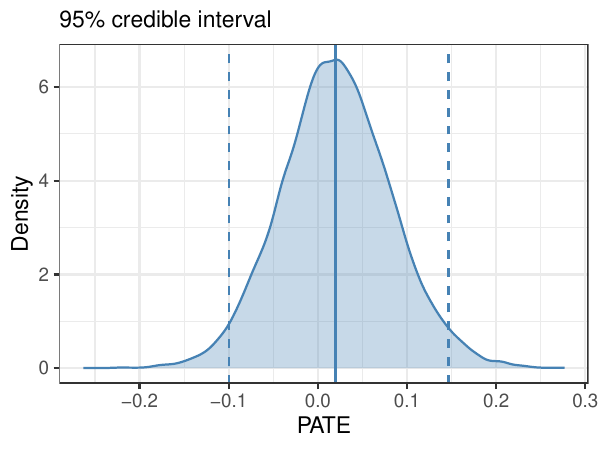} \includegraphics[width=0.48\linewidth]{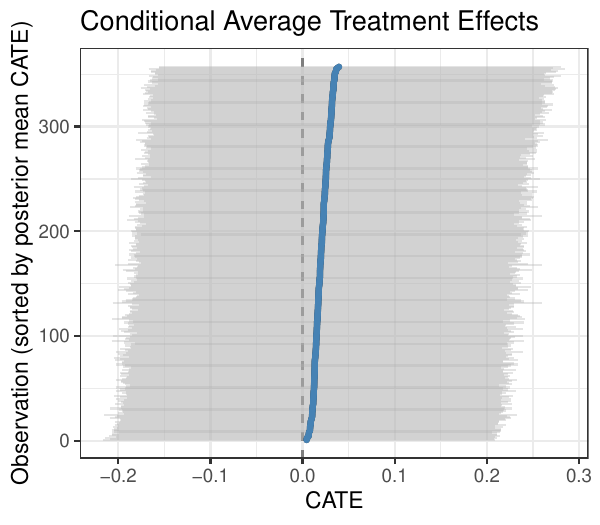} 

}

\caption[Treatment effect posteriors.]{Left: posterior density of the average treatment effect (ATE) of carboplatin versus cisplatin on the log-survival scale, with 95\% credible interval (dashed lines). Right: patient-level conditional average treatment effects (CATEs) sorted by posterior mean with 95\% credible intervals. The dashed line marks zero (no effect).}\label{fig:treatment-effects-fig}
\end{figure}

Figure \ref{fig:treatment-effects-fig} shows the treatment effect posteriors. The posterior mean ATE is \(0.021\) on the log-survival scale, corresponding to an approximate 2\% increase in survival time under carboplatin relative to cisplatin. The 95\% credible interval \([-0.10,\, 0.15]\) crosses zero, so the data do not provide clear evidence of a treatment effect. The individual CATEs cluster tightly around the ATE (CATE SD \(\approx 0.008\)). This suggests a relatively homogeneous treatment effect across the patient population.

Although the ovarian dataset contains right-censored observations, \CRANpkg{ShrinkageTrees} also supports interval-censored outcomes via \texttt{outcome\_type\ =\ "interval-censored"} with \texttt{left\_time} and \texttt{right\_time} arguments following the convention of \texttt{survival::Surv(type\ =\ "interval2")}. See \texttt{?ShrinkageTrees} for details and examples. We demonstrate this capability in the following simulation.

\subsection{Interval-censored outcomes: a simulation study}\label{interval-censored-outcomes-a-simulation-study}

We validate the interval-censored implementation of \CRANpkg{ShrinkageTrees} on a controlled simulation with known ground truth. The simulation compares the three priors available for interval-censored fits, since no other R package fits BART-style ensembles to interval-censored data. The comparison also guides the choice of prior when the dimension is high. The data-generating process combines the non-linear regression benchmark of \citet{friedman1991mars} with a sparse high-dimensional linear term. Covariates are \(X_{ij} \overset{\text{i.i.d.}}{\sim} \mathcal{U}[0, 1]\) and the noiseless mean function is:
\begin{equation}
f(x_i) = 10\, \sin(\pi\, x_{i1}\, x_{i2}) + 20\,(x_{i3} - 0.5)^2 + 10\, x_{i4} + 5\, x_{i5} + x_i^\top \beta,
\label{eq:friedman-f}
\end{equation}
where \(\beta_j \overset{\text{i.i.d.}}{\sim} (1 - s)\, \delta_0 + s\, \mathcal{N}(0, 1)\) with sparsity \(s = 0.05\). Within each replicate we draw a fresh \(\beta\), rescale \(f\) to unit empirical variance via a reference draw, and set \(\log T_i = f(x_i) + \varepsilon_i\) with \(\varepsilon_i \sim \mathcal{N}(0, \sigma^2)\). The noise scale \(\sigma\) is chosen so that \(\operatorname{var}(\log T_i) / \sigma^2 = 10/9\).

Each training subject is observed at three inspection times \(V_{i1} < V_{i2} < V_{i3}\) drawn from \(\text{Uniform}(L,\, U)\), with \(L\) a finite floor below the smallest realised \(\log T_i\) and \(U\) the empirical 95th percentile of \(\log T\). The recorded interval is \((V_{i,j-1},\, V_{ij}]\) for the smallest \(j\) with \(V_{ij} \ge \log T_i\), or \((V_{i3},\, \infty)\) if \(\log T_i\) exceeds the last inspection. A test set of size \(n_{\text{test}} = 1{,}000\) is used to evaluate predictions.

We fit three models per replicate on the same data: \texttt{SurvivalBART()}, \texttt{SurvivalDART()}, and \texttt{HorseTrees()} (the latter with \texttt{outcome\_type\ =\ "interval-censored"} and \texttt{k\ =\ 0.1}). All three use \(m = 200\) trees, and \(5{,}000\) burn-in and \(5{,}000\) posterior draws. We vary \(p \in \{50,\, 500,\, 5{,}000\}\) at fixed sample size \(n = 200\) and run 1,000 replicates per setting. For each fit, we report three metrics on both the training and test samples. We compare the posterior mean and pointwise 95\% credible intervals of the regression function against the true \(f\) via root mean squared error (RMSE), empirical coverage, and average interval length. The results are given in Table \ref{tab:sim-ic-table}. The Horseshoe Forest attains the lowest RMSE, the tightest credible intervals, and near-nominal 95\% coverage at all three \(p\) settings. The standard BART prior over-covers throughout, with intervals roughly 50--95\% wider than those of the Horseshoe Forest. The Dirichlet (DART) prior matches the nominal rate at \(p = 50\) but under-covers as \(p\) grows. Performance of the Horseshoe Forest is essentially flat across \(p\), scaling gracefully from the low-dimensional setting into the \(p \gg n\) regime.

\begin{table}[!h]
\centering
\caption{\label{tab:sim-ic-table}Interval-censored simulation: recovery of the regression function $f$,
       averaged over 1,000 replicates. Nominal coverage rate is 0.95.}
\centering
\begin{tabular}[t]{llrrrrrr}
\toprule
\multicolumn{2}{c}{ } & \multicolumn{3}{c}{Train} & \multicolumn{3}{c}{Test} \\
\cmidrule(l{3pt}r{3pt}){3-5} \cmidrule(l{3pt}r{3pt}){6-8}
$p$ & Method & RMSE & Coverage & Length & RMSE & Coverage & Length\\
\midrule
 & BART & 1.15 & 0.989 & 5.96 & 1.09 & 0.998 & 6.63\\

 & DART & 1.13 & 0.947 & 4.46 & 1.17 & 0.948 & 4.65\\

\multirow{-3}{*}{\raggedright\arraybackslash 50} & Horseshoe Forest & 0.95 & 0.963 & 3.92 & 0.98 & 0.963 & 4.05\\
\cmidrule{1-8}
 & BART & 1.31 & 0.979 & 6.46 & 1.07 & 1.000 & 7.64\\

 & DART & 1.26 & 0.924 & 4.57 & 1.32 & 0.923 & 4.80\\

\multirow{-3}{*}{\raggedright\arraybackslash 500} & Horseshoe Forest & 0.99 & 0.956 & 3.97 & 1.03 & 0.955 & 4.11\\
\cmidrule{1-8}
 & BART & 1.37 & 0.971 & 6.53 & 1.05 & 1.000 & 7.88\\

 & DART & 1.30 & 0.915 & 4.59 & 1.38 & 0.912 & 4.82\\

\multirow{-3}{*}{\raggedright\arraybackslash 5,000} & Horseshoe Forest & 1.00 & 0.951 & 3.99 & 1.03 & 0.951 & 4.13\\
\bottomrule
\end{tabular}
\end{table}

\section{Package architecture}\label{package-architecture}

\CRANpkg{ShrinkageTrees} is available on CRAN and on GitHub at \url{https://github.com/tijn-jacobs/ShrinkageTrees}. The R layer wraps a C++ backend through \CRANpkg{Rcpp} \citep{eddelbuettel2011rcpp}. We organise the description as follows: (i) fitting functions and S3 classes, (ii) the Rcpp backend, and (iii) workflow features such as multi-chain MCMC, \CRANpkg{coda} integration, and plotting. A \CRANpkg{testthat} \citep{wickham2011testthat} suite covers the fitting functions and user-facing methods. A vignette on CRAN demonstrates a full workflow.

\subsection{Fitting functions and S3 classes}\label{fitting-functions-and-s3-classes}

The package provides two main fitting functions, one for prediction and one for causal inference. It also provides six convenience wrappers that fix particular model configurations. The wrappers are named after the methods they fit (e.g., \texttt{HorseTrees()}, \texttt{SurvivalBART()}), so users who come with a method in mind can call it directly without configuring \texttt{prior\_type} or \texttt{outcome\_type} by hand. Table \ref{tab:fitting-functions} lists all eight functions. The main functions accept arguments that control the outcome type (\texttt{outcome\_type}, \texttt{timescale}), prior specification (\texttt{prior\_type}, \texttt{local\_hp}, \texttt{global\_hp}, and Dirichlet hyperparameters \texttt{a\_dirichlet}, \texttt{b\_dirichlet}, \texttt{rho\_dirichlet}), tree structure (\texttt{number\_of\_trees}, \texttt{power}, \texttt{base}, \texttt{p\_grow}, \texttt{p\_prune}), and MCMC settings (\texttt{N\_post}, \texttt{N\_burn}, \texttt{n\_chains}), among others. Each wrapper preselects a subset of these to expose a smaller, task-specific interface.

\begin{table}[!h]
\centering
\caption{\label{tab:fitting-functions}Fitting functions in \CRANpkg{ShrinkageTrees}. The first two
                        are the main functions. The remaining six are convenience wrappers
                        that fix particular model configurations.}
\centering
\begin{tabular}[t]{>{\raggedright\arraybackslash}p{4.5cm}>{\raggedright\arraybackslash}p{9.5cm}}
\toprule
Function & Description\\
\midrule
\texttt{ShrinkageTrees()} & Prediction model with user-specified step height prior (horseshoe, half-Cauchy, standard, Dirichlet)\\
\texttt{CausalShrinkageForest()} & $\tau$-learner causal model with user-specified step height priors for prognostic and treatment forests\\
\midrule
\texttt{HorseTrees()} & Prediction model with horseshoe prior, with a single tuning parameter $k$\\
\texttt{CausalHorseForest()} & $\tau$-learner causal model with horseshoe prior on both forests\\
\texttt{SurvivalBART()} & AFT survival with conjugate normal step height prior (standard BART)\\
\addlinespace
\texttt{SurvivalDART()} & AFT survival with Dirichlet splitting prior (DART)\\
\texttt{SurvivalBCF()} & AFT survival with $\tau$-learner decomposition and conjugate normal step height prior\\
\texttt{SurvivalShrinkageBCF()} & AFT survival with $\tau$-learner decomposition and horseshoe step height prior\\
\bottomrule
\end{tabular}
\end{table}

\texttt{ShrinkageTrees()} is the main prediction model function. It accepts a \texttt{prior\_type} argument that selects the step height prior: \texttt{"horseshoe"} (per-tree global scale), \texttt{"horseshoe\_fw"} (forest-wide global scale), \texttt{"half-cauchy"} (local shrinkage only), \texttt{"standard"} (conjugate normal, as in classical BART), or \texttt{"dirichlet"} (Dirichlet splitting, as in DART). \texttt{CausalShrinkageForest()} is the main causal-inference function. It mirrors this design for the \(\tau\)-learner with independent prior configurations for the prognostic and treatment effect forests. The six wrappers preselect a particular \texttt{prior\_type} and/or \texttt{outcome\_type} to give a smaller, task-specific interface: \texttt{HorseTrees()} and \texttt{CausalHorseForest()} fix the prior to the horseshoe and expose a single tuning parameter \(k\) controlling the half-Cauchy scales via \(\alpha_\tau = \alpha_\lambda = k / \sqrt{m}\). \texttt{SurvivalBART()} and \texttt{SurvivalDART()} fix the AFT survival outcome with the conjugate-normal and Dirichlet priors, respectively. \texttt{SurvivalBCF()} and \texttt{SurvivalShrinkageBCF()} do the same for the \(\tau\)-learner causal model.

All prediction model functions return an object of class \texttt{"ShrinkageTrees"} and all causal model functions return \texttt{"CausalShrinkageForest"}. Both classes provide \texttt{print()}, \texttt{summary()}, \texttt{predict()}, and \texttt{plot()} methods. The \texttt{summary()} method reports posterior summaries for \(\sigma\) and variable inclusion probabilities. For causal models, it also reports the ATE with a credible interval. The \texttt{predict()} method re-runs the MCMC sampler on the new covariate matrix using the training data and hyperparameters stored in the fitted object. It always generates posterior samples internally, regardless of whether \texttt{store\_posterior\_sample} was set during fitting. The returned \texttt{ShrinkageTreesPrediction} object contains posterior means and credible intervals. The \texttt{plot()} method produces MCMC diagnostics (traceplots and posterior densities for \(\sigma\)) and variable importance plots. For causal models, it also produces ATE and CATE visualisations. The plot methods use \CRANpkg{ggplot2} \citep{wickham2016ggplot2}, a suggested but not required dependency.

\subsection{Rcpp backend}\label{rcpp-backend}

We implement the computationally intensive components of the Gibbs sampler in C++ via \CRANpkg{Rcpp}. The C++ layer covers the Metropolis--Hastings updates for the tree topology (grow, prune, and change moves), the conditional updates for the step heights, the augmentation steps for the local and global scale parameters under horseshoe shrinkage, the update for \(\sigma\), and the data augmentation for censored event times. The R layer handles input validation, data preprocessing (centring, scaling, log-transformation of survival times), hyperparameter calibration, S3 class construction, and post-processing (back-transformation of predictions).

The C++ backend dispatches between two forest implementations depending on the step height prior. The \texttt{Forest} class implements reversible jump moves that update the tree and its step heights jointly. It is used for the \texttt{"horseshoe"}, \texttt{"horseshoe\_fw"}, and \texttt{"half-cauchy"} priors. The \texttt{StanForest} class implements moves that update the tree structure and the step heights sequentially. It is used for the \texttt{"standard"} and \texttt{"dirichlet"} priors. The latter approach is simpler and faster, but applies only when the step height prior is conjugate and global scale updates are unnecessary.

A single C++ class, \texttt{ScaleMixture}, holds the global--local step height prior and is queried by the tree updates whenever a new step height is proposed. The current implementations are \texttt{FixedVariance} (conjugate normal, used by BART and DART), \texttt{HalfCauchy} (local shrinkage only), \texttt{Horseshoe} (per-tree global--local shrinkage), and \texttt{Horseshoe\_fw} (forest-wide global shrinkage). The user's \texttt{prior\_type} argument selects the implementation at run time. Because the tree update and reversible jump routines interact with the prior only through the \texttt{ScaleMixture} interface, a developer can add a new global--local shrinkage prior by adding a new implementation and registering it. Examples include a regularised horseshoe or a Dirichlet--Laplace prior. The sampler logic does not change.

\begin{figure}

{\centering \includegraphics[width=0.94\linewidth]{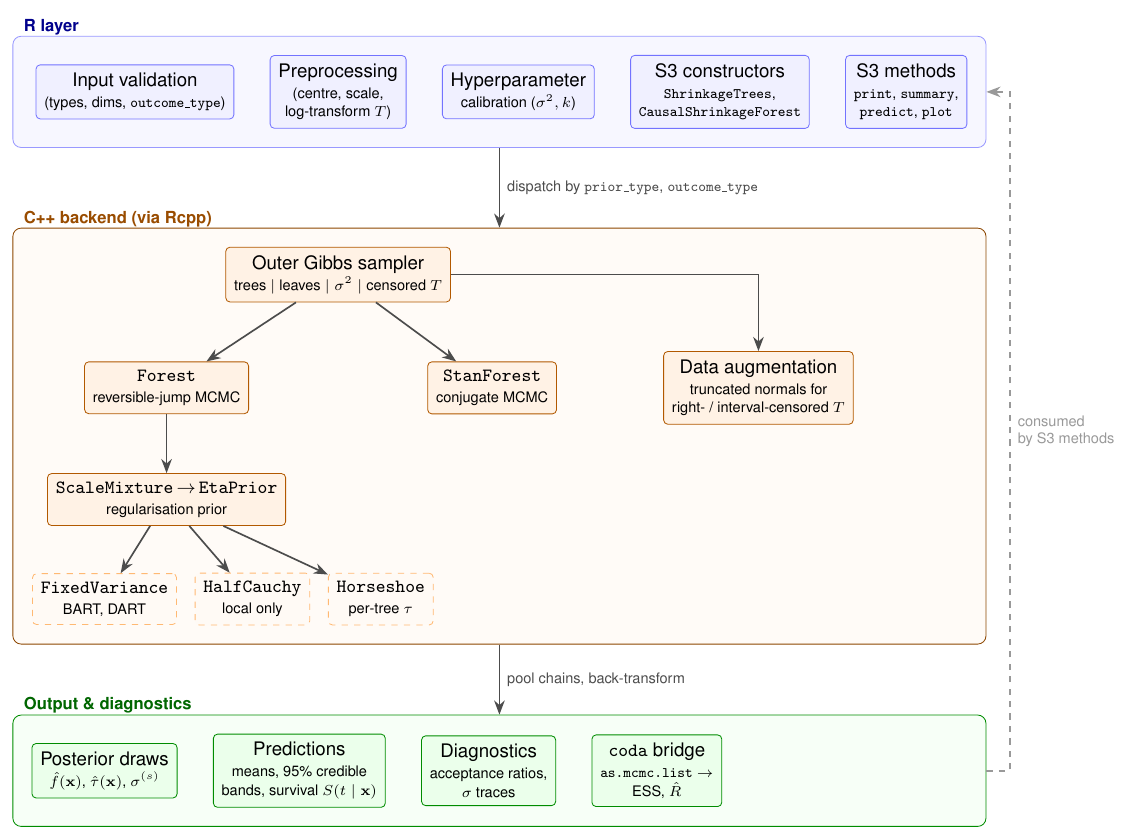} 

}

\caption[Package architecture.]{Package architecture. The R layer (top) handles input validation, preprocessing, hyperparameter calibration, S3 construction, and post-processing methods. The C++ backend (middle) runs the outer Gibbs sampler, dispatching to the reversible-jump \texttt{Forest} class or the conjugate birth--death \texttt{StanForest} class depending on the step height prior. The \texttt{ScaleMixture} wrapper selects one of four \texttt{EtaPrior} subclasses at runtime via a factory. The output layer (bottom) exposes posterior draws, predictions, diagnostics, and a \CRANpkg{coda} bridge to the user.}\label{fig:architecture}
\end{figure}

Figure \ref{fig:architecture} summarises the package architecture.

\subsection{Multi-chain MCMC}\label{multi-chain-mcmc}

All fitting functions accept an \texttt{n\_chains} argument. When \texttt{n\_chains\ \textgreater{}\ 1}, the package dispatches independent chains in parallel via \texttt{parallel::mclapply()}. On Windows, \texttt{mclapply()} falls back to sequential execution. The package pools the draws from all chains and recomputes posterior summaries; the returned object is a standard \texttt{"ShrinkageTrees"} or \texttt{"CausalShrinkageForest"} instance, so all downstream methods work without modification. Per-chain acceptance ratios and \(\sigma\) traces remain available for convergence assessment.

\subsection{Convergence diagnostics via coda}\label{convergence-diagnostics-via-coda}

We assess MCMC convergence via the \CRANpkg{coda} package \citep{plummer2006coda}. The package exposes posterior draws to \CRANpkg{coda} through an \texttt{as.mcmc.list()} S3 method. The method extracts the posterior draws of \(\sigma\), splits them by chain, wraps each chain in a \texttt{coda::mcmc} object, and returns a \texttt{coda::mcmc.list}. \CRANpkg{coda} is a suggested rather than required dependency; the package's other functionality does not depend on it. When \CRANpkg{coda} is installed, \texttt{summary()} reports the effective sample size (ESS) and, for multi-chain fits, the Gelman--Rubin \(\hat{R}\) statistic. Users who require more detailed diagnostics can pass the \texttt{mcmc.list} object directly to any \CRANpkg{coda} function, including \texttt{geweke.diag()}, \texttt{heidel.diag()}, \texttt{autocorr.plot()}, and \texttt{gelman.plot()}.

\subsection{Computational speed}\label{computational-speed}

We benchmark the wall-clock cost of fitting survival models in \CRANpkg{ShrinkageTrees} and compare it with \CRANpkg{BART}, the only other CRAN package that provides Bayesian tree ensembles for survival data. All timings are measured on a 2024 MacBook Air with an Apple M3 chip and 16 GB of memory, running R version 4.5.2 \citep{base}, \CRANpkg{ShrinkageTrees} 2.0.2, and \CRANpkg{BART} 2.9.10. Each fit uses \texttt{n\_chains\ =\ 4}, \(m = 200\) trees, and 1,000 posterior draws per chain.

We compare three fitting functions in \CRANpkg{ShrinkageTrees} against \texttt{abart()} from \CRANpkg{BART} on simulated right-censored survival data. The three are \texttt{SurvivalBART()}, \texttt{SurvivalDART()}, and \texttt{HorseTrees()}. \texttt{abart()} does not have an \texttt{n\_chains} argument. We therefore use its multi-chain wrapper \texttt{mc.abart()} with \texttt{mc.cores\ =\ 4}, which runs four independent chains in parallel, mirroring how \CRANpkg{ShrinkageTrees} dispatches its own chains.

\begin{figure}

{\centering \includegraphics[width=0.8\linewidth]{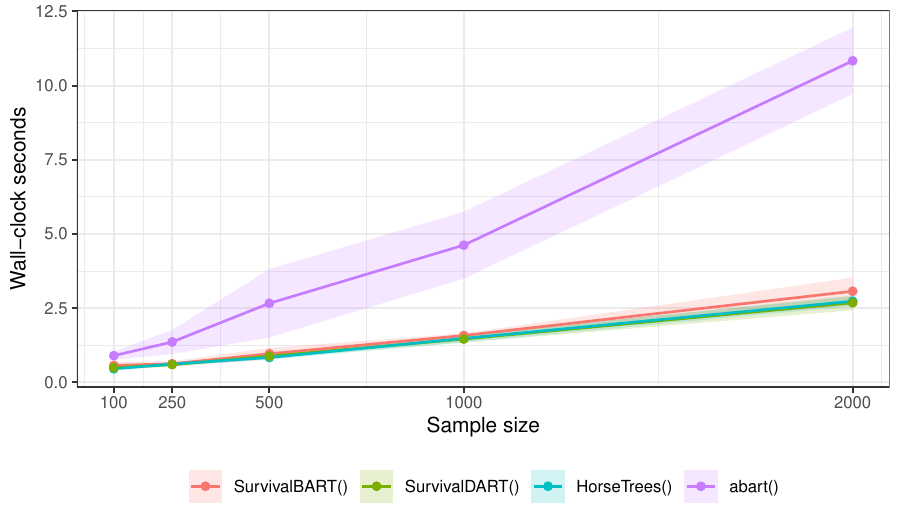} 

}

\caption[Computational scaling in $n$.]{Wall-clock seconds vs sample size $n$ for the four fitting functions, with $p = 100$ and $m = 200$ held fixed and $\mathtt{n\_chains} = 4$ throughout. Lines and points show means over three replications. Shaded ribbons show $\pm 1$ SD.}\label{fig:scaling}
\end{figure}

Figure \ref{fig:scaling} shows how wall-clock time scales with the sample size \(n\), with \(p = 100\) and \(m = 200\) fixed. Cost scales approximately linearly in \(n\) for all four fitting functions. The conjugate-prior models (\texttt{SurvivalBART()} and \texttt{SurvivalDART()}) are the closest comparators to \texttt{abart()} and run at similar cost. The horseshoe variant (\texttt{HorseTrees()}) is more expensive because it combines reversible jump tree updates with per-leaf local scale draws.

\section{Related packages}\label{related-packages}

Several R packages implement Bayesian tree ensembles. We review five comparator packages and contrast their capabilities with \CRANpkg{ShrinkageTrees}. Table \ref{tab:pkg-comparison} provides a summary.\footnote{Package descriptions reflect CRAN versions: BART 2.9.10, dbarts 0.9-33, bcf 2.0.2, stochtree 0.4.2, SoftBart 1.0.3.}

\CRANpkg{BART} \citep{sparapani2021nonparametric} is the most general-purpose package in this comparison and covers the widest range of outcome types. It supports continuous, binary, and multinomial outcomes, and is the only other CRAN package to offer Bayesian tree ensembles for survival data. Survival models include a discrete-time hazard formulation (\texttt{surv.bart()}), an AFT formulation (\texttt{abart()}), competing risks (\texttt{crisk.bart()}), and recurrent events (\texttt{recur.bart()}). The package also supports Dirichlet splitting priors via \texttt{sparse\ =\ TRUE}. \CRANpkg{BART} does not support global--local shrinkage priors on the step heights, causal forest models, or interval-censored survival outcomes. Multi-chain MCMC requires the prefixed \texttt{mc.*} wrappers (e.g.~\texttt{mc.abart()}).

\CRANpkg{dbarts} \citep{dorie2023dbarts} provides a BART sampler for continuous and binary outcomes. Its key feature is an updatable sampler object: predictors and responses can be modified mid-sampling, which makes \CRANpkg{dbarts} well suited as a conditional model within a larger Gibbs sampler. The package supports multi-chain MCMC but does not support survival models, \(\tau\)-learner models, or global--local shrinkage on the step heights.

\CRANpkg{bcf} \citep{hahn2020bayesian} implements the \(\tau\)-learner with separate prognostic and treatment-effect forests for continuous outcomes. It places half-Cauchy and half-Normal hyperpriors on the leaf scales of the two forests, respectively. The package does not support survival endpoints, global--local shrinkage priors, or Dirichlet splitting.

\CRANpkg{stochtree} \citep{stochtree} implements BART, XBART \citep{he2019xbart}, the \(\tau\)-learner, and XBCF for continuous, binary, and ordinal outcomes. It supports variance forests for heteroskedastic models and serialises fitted objects so they can be reused as warm starts. \CRANpkg{stochtree} does not support survival models, Dirichlet splitting priors, or global--local shrinkage on the step heights.

\CRANpkg{SoftBart} \citep{linero2018softbart} replaces the hard axis-aligned splits of standard BART with smooth logistic splitting rules, so the posterior mean is a continuous function of the covariates. Regularisation is achieved through a Dirichlet prior on the variable inclusion probabilities (the same mechanism as DART) rather than through priors on the step heights. This combination can improve performance when the true regression surface is smooth, but introduces additional computational cost. \CRANpkg{ShrinkageTrees} retains hard decision splits, as in classical BART, and instead places shrinkage priors on the step heights themselves. \CRANpkg{SoftBart} does not support survival endpoints, the \(\tau\)-learner, or global--local shrinkage on the step heights.

Across these packages, \CRANpkg{ShrinkageTrees} occupies a distinct niche: survival outcomes (right- and interval-censored), the \(\tau\)-learner for survival, and global--local shrinkage on the step heights.

\begin{table}[!h]
\centering
\caption{\label{tab:pkg-comparison}Comparison of Bayesian tree ensemble packages in R.}
\centering
\fontsize{9}{11}\selectfont
\begin{tabular}[t]{>{\raggedright\arraybackslash}p{2.2cm}>{\raggedright\arraybackslash}p{1.8cm}>{\raggedright\arraybackslash}p{1.8cm}>{\raggedright\arraybackslash}p{1.8cm}>{\raggedright\arraybackslash}p{1.8cm}>{\raggedright\arraybackslash}p{1.8cm}}
\toprule
Package & Survival & Interval\newline censoring & Causal\newline ($\tau$-learner) & Dirichlet\newline (DART) & Global--local\newline shrinkage\\
\midrule
BART & Yes & No & No & Yes & No\\
dbarts & No & No & No & No & No\\
bcf & No & No & Yes & No & No\\
stochtree & No & No & Yes & No & No\\
SoftBart & No & No & No & No & No\\
\addlinespace
ShrinkageTrees & Yes & Yes & Yes & Yes & Yes\\
\bottomrule
\end{tabular}
\end{table}

\section{Discussion and conclusion}\label{discussion-and-conclusion}

\CRANpkg{ShrinkageTrees} brings global--local shrinkage priors to Bayesian tree ensembles for survival analysis and causal inference. The package separates regularisation of the tree structure from regularisation of the step heights, so practitioners can tailor the regularisation strategy to the problem at hand. It is the first CRAN package to offer the \(\tau\)-learner for survival endpoints, horseshoe priors on tree ensemble step heights, and interval-censored survival outcomes within a single framework.

The package has several limitations that suggest directions for future work. All covariates must be numeric. The axis-aligned splitting rules handle continuous, binary, and ordered categorical variables (encoded as integers) directly, but the user must dummy-encode unordered categorical variables. This can be suboptimal for high-cardinality factors, where native subset splits would be more efficient. The current AFT formulation assumes normally distributed errors. Alternative parametric error distributions, such as the logistic or log-logistic, would broaden the package's applicability. Non-parametric error models, for example those built on stick-breaking process priors, are a further direction.

\CRANpkg{ShrinkageTrees} is available on CRAN and on GitHub at \url{https://github.com/tijn-jacobs/ShrinkageTrees}, where bug reports and feature requests can be filed via the issue tracker at \url{https://github.com/tijn-jacobs/ShrinkageTrees/issues}. \citet{jacobs2025horseshoe} develop the underlying methodology.

\section*{Acknowledgements}\label{acknowledgements}
\addcontentsline{toc}{section}{Acknowledgements}

The author thanks Wessel N. van Wieringen for carefully reading the manuscript and providing valuable feedback.

\section*{Funding}\label{funding}
\addcontentsline{toc}{section}{Funding}

Tijn Jacobs received funding from the European Research Council (ERC) under the European Union's Horizon Europe program under Grant agreement No.~101074082. Views and opinions expressed are however those of the author(s) only and do not necessarily reflect those of the European Union or the European Research Council Executive Agency. Neither the European Union nor the granting authority can be held responsible for them.

This work used the Dutch national e-infrastructure with the support of the
SURF Cooperative using grant no. EINF-11573.

\bibliography{RJreferences.bib}

@article{friedman1991mars,
  author  = {Jerome H. Friedman},
  title   = {Multivariate adaptive regression splines},
  journal = {The Annals of Statistics},
  year    = {1991},
  volume  = {19},
  number  = {1},
  pages   = {1--67},
  doi     = {10.1214/aos/1176347963}
}

@article{chipman1998bayesiancart,
  author  = {Hugh A. Chipman and Edward I. George and Robert E. McCulloch},
  title   = {{B}ayesian {CART} model search},
  journal = {Journal of the American Statistical Association},
  year    = {1998},
  volume  = {93},
  number  = {443},
  pages   = {935--948},
  doi     = {10.1080/01621459.1998.10473750}
}

@inproceedings{rockova2020theory,
  author    = {Ro\v{c}kov\'a, Veronika and Saha, Enakshi},
  title     = {On theory for {BART}},
  booktitle = {Proceedings of the Twenty-Second International Conference on Artificial Intelligence and Statistics},
  pages     = {2839--2848},
  year      = {2019},
  editor    = {Chaudhuri, Kamalika and Sugiyama, Masashi},
  volume    = {89},
  series    = {Proceedings of Machine Learning Research},
  month     = {16--18 Apr},
  publisher = {PMLR},
  url       = {https://proceedings.mlr.press/v89/rockova19a.html}
}

@article{chipman2010bart,
  author  = {Hugh A. Chipman and Edward I. George and Robert E. McCulloch},
  title   = {{BART}: {B}ayesian additive regression trees},
  journal = {The Annals of Applied Statistics},
  year    = {2010},
  volume  = {4},
  number  = {1},
  pages   = {266--298},
  doi     = {10.1214/09-AOAS285}
}

@article{hastie2000backfitting,
  author  = {Trevor Hastie and Robert Tibshirani},
  title   = {{B}ayesian backfitting (with comments and a rejoinder by the authors)},
  journal = {Statistical Science},
  year    = {2000},
  volume  = {15},
  number  = {3},
  pages   = {196--223},
  doi     = {10.1214/ss/1009212815}
}

@article{carvalho2010horseshoe,
  author  = {Carlos M. Carvalho and Nicholas G. Polson and James G. Scott},
  title   = {The horseshoe estimator for sparse signals},
  journal = {Biometrika},
  year    = {2010},
  volume  = {97},
  number  = {2},
  pages   = {465--480},
  doi     = {10.1093/biomet/asq017}
}

@article{jacobs2025horseshoe,
  author  = {Tijn Jacobs and Wessel N. van Wieringen and St\'{e}phanie L. van der Pas},
  title   = {Horseshoe forests for high-dimensional causal survival analysis},
  journal = {arXiv preprint arXiv:2507.22004},
  year    = {2025},
  note    = {Accepted for publication in \emph{Bayesian Analysis}},
  eprint  = {2507.22004},
  archiveprefix = {arXiv},
  primaryclass  = {stat.ME},
  doi     = {10.48550/arXiv.2507.22004}
}

@article{linero2018dart,
  author  = {Antonio R. Linero},
  title   = {{B}ayesian regression trees for high-dimensional prediction and variable selection},
  journal = {Journal of the American Statistical Association},
  year    = {2018},
  volume  = {113},
  number  = {522},
  pages   = {626--636},
  doi     = {10.1080/01621459.2016.1264957}
}

@article{hahn2020bayesian,
  author  = {P. Richard Hahn and Jared S. Murray and Carlos M. Carvalho},
  title   = {{B}ayesian regression tree models for causal inference: Regularization, confounding, and heterogeneous effects},
  journal = {Bayesian Analysis},
  year    = {2020},
  volume  = {15},
  number  = {3},
  pages   = {965--1056},
  doi     = {10.1214/19-BA1195}
}

@article{sparapani2021nonparametric,
  author  = {Rodney Sparapani and Charles Spanbauer and Robert McCulloch},
  title   = {Nonparametric machine learning and efficient computation with {B}ayesian additive regression trees: The {BART} {R} package},
  journal = {Journal of Statistical Software},
  year    = {2021},
  volume  = {97},
  number  = {1},
  pages   = {1--66},
  doi     = {10.18637/jss.v097.i01}
}

@Manual{dorie2023dbarts,
  title  = {{dbarts}: Discrete {B}ayesian additive regression trees sampler},
  author = {Vincent Dorie and Hugh Chipman and Robert McCulloch},
  year   = {2026},
  note   = {R package version 0.9-33},
  url    = {https://CRAN.R-project.org/package=dbarts},
  doi    = {10.32614/CRAN.package.dbarts}
}

@Manual{stochtree,
  title  = {{stochtree}: Stochastic tree ensembles ({XBART} and {BART}) for supervised learning and causal inference},
  author = {Drew Herren and Richard Hahn and Jared Murray and Carlos Carvalho and Jingyu He},
  year   = {2026},
  note   = {R package version 0.4.2},
  url    = {https://CRAN.R-project.org/package=stochtree},
  doi    = {10.32614/CRAN.package.stochtree}
}

@article{kapelner2016bartmachine,
  author  = {Adam Kapelner and Justin Bleich},
  title   = {{bartMachine}: Machine learning with {B}ayesian additive regression trees},
  journal = {Journal of Statistical Software},
  year    = {2016},
  volume  = {70},
  number  = {4},
  pages   = {1--40},
  doi     = {10.18637/jss.v070.i04}
}

@article{linero2018softbart,
  author  = {Antonio R. Linero and Yun Yang},
  title   = {{B}ayesian regression tree ensembles that adapt to smoothness and sparsity},
  journal = {Journal of the Royal Statistical Society: Series B (Statistical Methodology)},
  year    = {2018},
  volume  = {80},
  number  = {5},
  pages   = {1087--1110},
  doi     = {10.1111/rssb.12293}
}

@article{polson2012halfcauchy,
  author  = {Nicholas G. Polson and James G. Scott},
  title   = {On the half-{C}auchy prior for a global scale parameter},
  journal = {Bayesian Analysis},
  year    = {2012},
  volume  = {7},
  number  = {4},
  pages   = {887--902},
  doi     = {10.1214/12-BA730}
}

@article{plummer2006coda,
  author  = {Martyn Plummer and Nicky Best and Kate Cowles and Karen Vines},
  title   = {{CODA}: Convergence diagnosis and output analysis for {MCMC}},
  journal = {R News},
  year    = {2006},
  volume  = {6},
  number  = {1},
  pages   = {7--11},
  url     = {https://journal.r-project.org/archive/}
}

@article{caron2022estimating,
  author  = {Alberto Caron and Gianluca Baio and Ioanna Manolopoulou},
  title   = {Estimating individual treatment effects using non-parametric regression models: A review},
  journal = {Journal of the Royal Statistical Society Series A: Statistics in Society},
  year    = {2022},
  volume  = {185},
  number  = {3},
  pages   = {1115--1149},
  doi     = {10.1111/rssa.12824}
}

@article{tcga2011ovarian,
  author  = {{The Cancer Genome Atlas Research Network}},
  title   = {Integrated genomic analyses of ovarian carcinoma},
  journal = {Nature},
  year    = {2011},
  volume  = {474},
  pages   = {609--615},
  doi     = {10.1038/nature10166}
}

@article{harrell1996concordance,
  author  = {Frank E. Harrell and Kerry L. Lee and Daniel B. Mark},
  title   = {Multivariable prognostic models: Issues in developing models, evaluating assumptions and adequacy, and measuring and reducing errors},
  journal = {Statistics in Medicine},
  year    = {1996},
  volume  = {15},
  number  = {4},
  pages   = {361--387},
  doi     = {10.1002/(SICI)1097-0258(19960229)15:4<361::AID-SIM168>3.0.CO;2-4}
}

@article{eddelbuettel2011rcpp,
  author  = {Dirk Eddelbuettel and Romain Fran\c{c}ois},
  title   = {{Rcpp}: Seamless {R} and {C++} integration},
  journal = {Journal of Statistical Software},
  year    = {2011},
  volume  = {40},
  number  = {8},
  pages   = {1--18},
  doi     = {10.18637/jss.v040.i08}
}

@article{hill2011bayesian,
  author  = {Hill, Jennifer L},
  title   = {{B}ayesian nonparametric modeling for causal inference},
  journal = {Journal of Computational and Graphical Statistics},
  volume  = {20},
  number  = {1},
  pages   = {217--240},
  year    = {2011},
  doi     = {10.1198/jcgs.2010.08162}
}

@article{dorie2019automated,
  author  = {Dorie, Vincent and Hill, Jennifer and Shalit, Uri and Scott, Marc and Cervone, Dan},
  title   = {Automated versus do-it-yourself methods for causal inference: Lessons learned from a data analysis competition},
  journal = {Statistical Science},
  volume  = {34},
  number  = {1},
  pages   = {43--68},
  year    = {2019},
  doi     = {10.1214/18-STS667}
}

@article{li2023bayesian,
  author  = {Li, Fan and Ding, Peng and Mealli, Fabrizia},
  title   = {{B}ayesian causal inference: A critical review},
  journal = {Philosophical Transactions of the Royal Society A},
  volume  = {381},
  number  = {2247},
  pages   = {20220153},
  year    = {2023},
  doi     = {10.1098/rsta.2022.0153}
}

@article{rubin1981bayesian,
  author  = {Rubin, Donald B},
  title   = {The {B}ayesian bootstrap},
  journal = {The Annals of Statistics},
  volume  = {9},
  number  = {1},
  pages   = {130--134},
  year    = {1981},
  doi     = {10.1214/aos/1176345338}
}

@inproceedings{he2019xbart,
  author    = {He, Jingyu and Yalov, Sergiy and Hahn, P. Richard},
  title     = {{XBART}: Accelerated {B}ayesian additive regression trees},
  booktitle = {Proceedings of the 22nd International Conference on Artificial Intelligence and Statistics (AISTATS)},
  volume    = {89},
  pages     = {1130--1138},
  year      = {2019},
  url       = {https://proceedings.mlr.press/v89/he19a.html}
}

@Manual{therneau2024survival,
  title  = {A package for survival analysis in {R}},
  author = {Terry M. Therneau},
  year   = {2024},
  note   = {R package version 3.8-3},
  url    = {https://CRAN.R-project.org/package=survival}
}

@Book{wickham2016ggplot2,
  author    = {Hadley Wickham},
  title     = {{ggplot2}: Elegant graphics for data analysis},
  publisher = {Springer-Verlag New York},
  year      = {2016},
  isbn      = {978-3-319-24277-4},
  url       = {https://ggplot2.tidyverse.org}
}

@Article{wickham2011testthat,
  author  = {Hadley Wickham},
  title   = {{testthat}: Get started with testing},
  journal = {The R Journal},
  year    = {2011},
  volume  = {3},
  pages   = {5--10},
  url     = {https://journal.r-project.org/articles/RJ-2011-002/}
}

@Manual{base,
  title        = {{R}: A language and environment for statistical computing},
  author       = {{R Core Team}},
  organization = {R Foundation for Statistical Computing},
  address      = {Vienna, Austria},
  year         = {2025},
  url          = {https://www.R-project.org/}
}

\end{article}

\end{document}